\documentclass[%        Class options:
aps,%                   American Physical Society
prd,%                   Physical Review D
showpacs,%              Displays PACS after abstract
preprint,%              Preprint layout
tightenlines,%          Single spaced lines
nofootinbib,%           Does not treat footnotes as references
floatfix]%              Fixes float errors
{revtex4}%              REVTEX 4 Package used
\usepackage{graphicx,%  Default Latex 2eps package for embedding figures
longtable%             Useful for long table
}

\begin{document}
\title{Probing non-standard decoherence effects\\
with solar and KamLAND neutrinos}
\author{G.L.~Fogli$^1$, E.~Lisi$^1$, A.~Marrone$^1$,
D.~Montanino$^2$, and A.~Palazzo$^{3,1}$}
\address{
$^1$ Dipartimento di Fisica and Sezione INFN
di Bari, Via Amendola 173, 70126, Bari, Italy\\
$^2$ Dipartimento di Fisica and Sezione INFN di Lecce\\
Via Arnesano, 73100 Lecce, Italy\\
$^3$ Astrophysics, Denys Wilkinson Building, Keble Road, OX1 3RH,
Oxford, United Kingdom\\}

\begin{abstract}

It has been speculated that quantum gravity might induce a
``foamy'' space-time structure at small scales, randomly
perturbing the propagation phases of free-streaming particles
(such as kaons, neutrons, or neutrinos). Particle interferometry
might then reveal non-standard decoherence effects, in addition to
standard ones (due to, e.g., finite source size and detector
resolution.) In this work we discuss the phenomenology of such
non-standard effects in the propagation of electron neutrinos in
the Sun and in the long-baseline reactor experiment KamLAND, which
jointly provide us with the best available probes of decoherence
at neutrino energies $E\sim$~few~MeV. In the solar neutrino case,
by means of a perturbative approach, decoherence is shown to
modify the standard (adiabatic) propagation in matter through a
calculable damping factor. By assuming a power-law dependence of
decoherence effects in the energy domain ($E^n$ with
$n=0,\,\pm1,\,\pm2$), theoretical predictions for two-family
neutrino mixing are compared with the data and discussed. We find
that neither solar nor KamLAND data show evidence in favor of
non-standard decoherence effects, whose characteristic parameter
$\gamma_0$ can thus be significantly constrained. In the
``Lorentz-invariant'' case $n=-1$, we obtain the upper limit
$\gamma_0<0.78\times 10^{-26}$ GeV at $95\%$ C.L. In the specific
case $n=-2$, the constraints can also be interpreted as bounds on
possible matter density fluctuations in the Sun, which we improve
by a factor of $\sim 2$ with respect to previous analyses.
\end{abstract}

\pacs{14.60.Pq, 26.65.+t, 03.65.Yz, 04.60.-m}
\maketitle

\section{Introduction}

Although a satisfactory theory of quantum gravity is still
elusive, it has been speculated that it should eventually entail
violations of basic quantum mechanics, including the spontaneous
evolution of pure states into mixed (decoherent) states
\cite{Haw75} through unavoidable interactions with a pervasive and
``foamy'' space-time fabric at the Planck scale~\cite{Foam}. The
pioneering paper~\cite{El84} showed that such hypothetical source
of decoherence might become manifest in oscillating systems which
propagate over macroscopical distances, through additional
smearing effects in the observable interferometric pattern
(besides the usual smearing effects due, e.g., to the finite
source size and the detector resolution). However, lacking an ``ab
initio'' theory of quantum gravity decoherence, its effects can
only be parameterized in a model-dependent (and somewhat
arbitrary) way. Searches with neutral kaon
oscillations~\cite{El84,Banks84,kaon,Bern06}, neutron
interferometry~\cite{El84,neut} and, more recently, neutrino
oscillations~\cite{sol1,sol2,atm1,Lisi1,Lisi2,Gago01,Blenn}, have
found no evidence for such effects so far, and have placed bounds
on model parameters.

Quantum gravity effects in neutrino systems have been investigated
with increasing attention in the last decade, as a result of the
evidence for neutrino flavor oscillations. Early attempts
tried to interpret the solar neutrino puzzle~\cite{sol1,sol2,atm1}
or the atmospheric $\nu$ anomaly~\cite{atm1} in terms of decoherence.
After the first convincing evidence for
atmospheric neutrino oscillations~\cite{Atm_evid}, a quantitative
analysis was performed in~\cite{Lisi1}, considering possible
decoherence effects in the $\nu_\mu \to \nu_\mu$ channel (see
also~\cite{Lisi2}). The phenomenology of terrestrial neutrino
experiments has also been investigated in~\cite{Gago01,Blenn}. More
recently, prospective studies have focused on decoherence effects
in high energy neutrinos~\cite{Ahlu01,Hoop1_04,Hoop1_05,Hoop2_05,Morg_06},
observable in next generation neutrino telescopes. Furthermore, more
formal aspects of quantum gravity decoherence in neutrino
systems have been developed~\cite{Bena00,Bena01,Bare1_05,Bare2_05,Bare3_05,Mavro_05,Mavro_06}.
This is only a fraction of the related literature, which testifies the wide
and increasing interest in the subject.

Despite this interest, to our knowledge such decoherence effects
have not been systematically investigated in the light of the
solar neutrino experiments performed in the last few years. Solar
neutrino oscillations~\cite{Pont} dominated by matter effects
\cite{Matt,Adia} are currently well established by solar neutrino
experiments~\cite{CHL,Kam,SAGE,GALL,GGNO,Gavr,SKso,SK04,SNO1,SNO2,SNOL}
and have been independently confirmed by the long-baseline reactor
experiment KamLAND~\cite{Kam1,Kam2}. The striking agreement
between solar and KamLAND results determines a unique solution in
the mass-mixing parameter space [the so-called Large Mixing Angle
(LMA) solution, see e.g.~\cite{Rev,Rev_val}], provides indirect
evidence for matter effects with standard amplitude~\cite{gf04},
and generally (although not always \cite{Robust}) implies that
additional, non-standard physics effects may play only a
subleading role, if any. In particular, the KamLAND collaboration
has exploited the observation of half oscillation cycle in the
energy spectrum~\cite{Kam2} to exclude decoherence as a dominant
explanation of their data.

The main purpose of this paper is then to study decoherence as a
{\em subdominant\/} effect in solar and KamLAND neutrino
oscillations. Modifications of the standard oscillation formulae
in the presence of decoherence, and qualitative bounds on
decoherence parameters, are discussed in Sec.~II and III for
KamLAND and solar neutrinos, respectively. Quantitative bounds on
subdominant decoherence effects from a joint analysis of solar and
KamLAND data are studied in Sec.~IV. Implications for decoherence
induced by matter fluctuations in the Sun are discussed in Sec.~V.
The main results are finally summarized in Sec.~VI. The solar
neutrino flavor evolution in the presence of standard matter
effects plus non-standard decoherence is discussed in a technical
Appendix.

\section{Oscillations with(out) decoherence in KamLAND}

Here and in the following, we assume the standard notation \cite{PDG} for
neutrino mixing, and set the small mixing angle
$\theta_{13}$ to zero for the sake of simplicity.
For $\theta_{13}=0$, oscillations in the
 $\nu_e\to\nu_e$ channel probed by long-baseline reactor (KamLAND) and by
solar neutrinos are driven by only two parameters: the mixing
angle $\theta_{12}$ and the neutrino
squared mass difference $\delta m^2=m^2_2-m^2_1$. In particular,
the standard $\nu_e$ survival probability over a baseline $L$
in KamLAND reads:
%.......................................................................
\begin{equation}
P_{ee}   = 1-\frac{1}{2}\sin^2 2\theta_{12}
        \left(1-\cos \left(\frac{\delta m^2 L}{2E}\right)\right)\,.
\label{K_osc}
\end{equation}
%.......................................................................

In the presence of additional decoherence effects, the oscillating
factor is exponentially suppressed, as shown in \cite{Lisi1} for
the atmospheric $\nu_\mu\to\nu_\mu$ channel. By changing the
appropriate parameters for the KamLAND $\nu_e\to\nu_e$ channel,
the results of \cite{Lisi1} lead to the following modification of
the previous equation,
%.......................................................................
\begin{equation}
P_{ee}   = 1-\frac{1}{2}\sin^2 2\theta_{12}
        \left(1-e^{-\gamma L}\cos \left(\frac{\delta m^2 L}{2E}\right)\right)\,,
\label{K_osc_dec}
\end{equation}
%.......................................................................
where the dimensional parameter $\gamma$ represents the inverse of
the decoherence length after which the neutrino system gets mixed.%
%-------------------------------------------------
\footnote{Units: $[\gamma]=$~1/length~=~energy. Conversion factor:
(1~km)$^{-1}=1.97\times 10^{-19}$~GeV.}
%---------------------------------------------------
Equation~(\ref{K_osc_dec}) includes the limiting
cases of pure oscillations ($\gamma=0$ and
$\delta m^2\neq 0$)
and of pure decoherence ($\gamma\neq 0$ and $\delta m^2=0$).

Unfortunately, lacking a fundamental theory for quantum gravity,
the dependence of $\gamma$ on the underlying dynamical and
kinematical parameters (most notably the neutrino energy $E$) is
unknown. Following common practice, such ignorance is
parameterized in a power-law form
%%%%%%%%%%%%%%%%%%%%%%%%%%%%%%%%%%%%%%%%%%%%%%%%%%%%%%%%%%%%%%%%%%%%
\begin{equation}
\gamma = \gamma_0 \left(\frac{E}{E_0}\right)^n ,\label{eq:gamma}
\end{equation}
%%%%%%%%%%%%%%%%%%%%%%%%%%%%%%%%%%%%%%%%%%%%%%%%%%%%%%%%%%%%%%%%%%%%
where $E_0$ is an arbitrary pivot energy scale, which we set as
$E_0=1$~GeV in order to facilitate the comparison with limits on
$\gamma$-parameters investigated in other contexts (as reviewed,
e.g., in \cite{Ancho}). We shall consider only five possible
integer exponents,
%.................................................................
\begin{equation}
n=0,\,\pm1,\,\pm2\,,
\label{exponents}
\end{equation}
%.................................................................
which include the following cases of interest: The ``energy
independent'' case ($n = 0$); the ``Lorentz invariant''
case~\cite{Lisi1} ($n= -1$); the case $n=+2$ that can arise in
some D-brane or quantum-gravity models, in which $\gamma_0\sim
O(E_0^2/M_{\rm Planck})\sim 10^{-19}$~GeV is expected (see,
e.g.~\cite{string}); and the case where decoherence might be
induced by ``matter density fluctuations'' rather than by quantum
gravity ($n=-2$, see Sec.~V).

As previously remarked, the KamLAND collaboration \cite{Kam2} (see
also~\cite{Schwetz}) has ruled out pure decoherence in the Lorentz invariant case ($n=-1$). In our statistical $\chi^2$ analysis,
we also find that this case is rejected at $3.6\sigma$
(i.e., $\Delta \chi^2=13$ with respect to pure oscillations). In addition,
we find that the other exponents in Eq.~(\ref{exponents}) are also
rejected at $>3\sigma$ for the
pure decoherence case. Therefore,
decoherence effects can only be subdominant in KamLAND, namely
%.................................................................
\begin{equation}
\gamma L \ll 1\ .
\label{gless}
\end{equation}
%.................................................................
For typical KamLAND neutrino energies $(E\sim \mathrm{few\ MeV})$
and baselines $(L\sim 2\times 10^2$~km), the above inequality
implies upper bounds on $\gamma_0$, which range from $\gamma_0\ll
10^{-26}$~GeV ($n=-2$) to $\gamma_0\ll 10^{-16}$~GeV ($n=+2$). We
do not refine the analysis of such bounds (placed by KamLAND
alone), since they are superseded by solar data constraints, as
shown in the next Section.

\section{Oscillations with(out) decoherence effects in solar neutrinos}

The survival probability describing standard adiabatic
 $\nu_e$ transitions in the solar matter is given by
the simple formula (up to small Earth matter effects)
%%%%%%%%%%%%%%%%%%%%%%%%%%%%%%%%%%%%%%%%%%%%%%%%%%%%%%%%%%%%%%%%%%
\begin{equation}
P_{ee}^\odot=\frac{1}{2}\left(1+ \cos 2\tilde\theta_{12}(r_0)
\cos 2\theta_{12}\right) \,,
\label{sol_osc}
\end{equation}
%%%%%%%%%%%%%%%%%%%%%%%%%%%%%%%%%%%%%%%%%%%%%%%%%%%%%%%%%%%%%%%%%%
where $\tilde \theta_{12}(r_0)$ is the energy-dependent
effective mixing angle in matter
at the production radius $r_0$
 (see, e.g., \cite{Adiabat} and references therein).

In the presence of non-standard decoherence effects, we find that
the energy dependent term is modulated by an exponential factor,
%%%%%%%%%%%%%%%%%%%%%%%%%%%%%%%%%%%%%%%%%%%%%%%%%%%%%%%%%%%%%%%%%%
\begin{equation}
P_{ee}^\odot=\frac{1}{2}\left(1+ e^{-\gamma_\odot R_\odot}\cos 2\tilde\theta_{12}(r_0)
\cos 2\theta_{12}\right) \,,
\label{sol_osc_dec}
\end{equation}
%%%%%%%%%%%%%%%%%%%%%%%%%%%%%%%%%%%%%%%%%%%%%%%%%%%%%%%%%%%%%%%%%%
where $R_\odot=6.96\times 10^{5}~\mathrm{km}$ is the Sun radius, while
$\gamma_\odot$ is defined as
%.............................................................
\begin{equation}
\gamma_\odot = \gamma_0\,g_n(E)\ ,
\label{gamma_sol}
\end{equation}
%...............................................................
where the dimensionless function $g_n(E)$ embeds, besides the
power-law dependence $E^n$, also the information about the solar
density profile [which is instead absent in Eq.~(6)]. The reader
is referred to the Appendix for a derivation of
Eq.~(\ref{sol_osc_dec}) and for details about the function
$g_n(E)$.

Equation~(\ref{sol_osc_dec}) includes the subcase of pure
oscillations ($\gamma_0=0$ and $\delta m^2\neq 0$), but not the
subcase of pure decoherence ($\gamma_0\neq 0$ and $\delta m^2=0$),
since the limit $\delta m^2\to 0$ would entail strongly
nonadiabatic transitions, and thus a breakdown of the adiabatic
approximation assumed above. However, as noted in the previous
section, the KamLAND data exclude the limit $\delta m^2\to 0$;
moreover, they require $\delta m^2$ values which are high enough
to guarantee the validity of the adiabatic approximations, with or
without subleading decoherence effects (as we have numerically
verified). Therefore, for solar neutrinos, the only
phenomenologically
relevant cases are those including oscillations plus decoherence.%
%........................................................
\footnote{For the sake of curiosity, we have anyway calculated
$P_{ee}^\odot$ for the pure decoherence case, by numerically
solving the neutrino evolution equations (discussed in the
Appendix) for $\delta m^2=0$ and $\gamma \neq 0$. We always find
$P_{ee}^\odot>1/2$, which is forbidden by $^8$B solar neutrino
data \protect\cite{gf04}.}
%.........................................................

We have analyzed all the available solar neutrino data with
$(\delta m^2,\theta_{12},\gamma_0)$ taken as free parameters. It
turns out that, despite the allowance for an extra degree of
freedom ($\gamma_0$), the data always prefer the pure oscillations
($\gamma_0=0$) as best fit, independently of the power-law index
in Eq.~(\ref{exponents}). Since there are no indications in favor
of decoherence effects, the exponent in Eq.~(\ref{sol_osc_dec}) is
expected to be small,
%.................................................................
\begin{equation}
\gamma_\odot R_\odot \ll 1\ .
\label{g_sol_less}
\end{equation}
%.................................................................
For typical solar neutrino energies $E\sim 10~\mathrm{MeV}$, it turns out that
 $g_n(E)\sim 0.2\times 10^{-2n}$ (see the Appendix), and
the above inequality can be translated into upper bounds on
$\gamma_0$, which range from $\gamma_0\ll 10^{-28}$~GeV (for
$n=-2$) to $\gamma_0\ll 10^{-20}$~GeV (for $n=+2$). Such bounds
are two to four orders of magnitude stronger than those placed by
KamLAND alone (see the end of the previous Section). A useful
complementarity then emerges between solar and KamLAND data in
joint fits: The former dominate the constraints on decoherence
effects, while the latter fix the mass-mixing parameters
independently of (negligible) decoherence effects.

\section{Combination of Solar and KamLAND data: Results and discussion}

We have performed a joint analysis of solar and KamLAND data%
%-------------
\footnote{The details of the data set, of the solar model used
\protect\cite{BSOP}
and of the statistical $\chi^2$ analysis have been reported in
\protect\cite{Rev} and are not
repeated here.}
%...................
 in the
$(\delta m^2,\,\sin^2\theta_{12},\,\gamma_0)$ parameter space for
the five power-law exponents $n=0,\,\pm1,\,\pm2$. The main results
are: (i) $\gamma_0=0$ is always preferred at best fit, i.e., there
is no indication in favor of decoherence effects; (ii) the best
fit values and the marginalized bounds for $(\delta
m^2,\sin^2\theta_{12})$ do not appreciably change from those
obtained in the pure oscillation case, namely, $\delta
m^2=(7.92\pm 0.71)\times 10^{-5}$ eV$^2$ and
$\sin^2\theta_{12}=(0.314^{+0.057}_{-0.047})$ at $\pm2\sigma$
\cite{Rev}; (iii) significant upper bounds can be set on the
decoherence parameter $\gamma_0$. Our limits on $\gamma_0$ are
given numerically in Table~I (at the $2\sigma$ level,
$\Delta\chi^2=4$) and graphically in Fig.~1 (at $2\sigma$ and
$3\sigma$ level). Such limits are consistent with those discussed
qualitatively after Eq.~(\ref{g_sol_less}) in the previous
Section.

%===========================================================================
\begin{table}[t]
\caption{\label{table:limits} Upper limits on the decoherence
parameter $\gamma_0$ obtained for different values of $n$ from a global fit
to solar and KamLAND data, after marginalization of the mass-mixing
parameters. The limits refer to 95\% C.L.\ (i.e., $2\sigma$, or $\Delta\chi^2=4$).}
\begin{ruledtabular}
\begin{tabular}{rc}
~~$n$        & $\gamma_0 ~(\mathrm{GeV}$)    \\[4pt]
\hline%---------------------------------------------------------------------
$-2$    &       $<0.81 \times 10^{-28}$           \\
$-1$    &       $<0.78 \times 10^{-26}$           \\
$0$     &       $<0.67 \times 10^{-24}$           \\
$+1$    &       $<0.58 \times 10^{-22}$           \\
$+2$    &       $<0.47 \times 10^{-20}$
\end{tabular}
\end{ruledtabular}
\end{table}
%============================================================================

Figure~1 clearly shows that the bounds on $\gamma_0$ scale with
$n$ almost exactly as a power law, changing by about two decades
for $|\Delta n|=1$. The reason is that the bounds are dominated by
solar neutrino data, and in particular by data probing the $^8$B
neutrinos in a relatively narrow energy range around $E\sim
10$~MeV; the power-law dependence assumed in Eq.~(\ref{eq:gamma})
and embedded in the function $g_n(E)$ then implies that the
parameter $\gamma_0$ scales roughly as $(E/E_0)^{-n}\sim 10^{2n}$.

Although the case of no decoherence ($\gamma_0=0$)
is preferred, it makes sense
to ask what one should observe for decoherence effects as large as
currently allowed by the data
at, say, the $2\sigma$ level.
Figure~2 compares the
$P_{ee}^\odot$ energy profile for the cases of pure
oscillations (left panel) and of oscillations plus decoherence
(right panel), where $\gamma_0$ is taken equal to the upper bound at
$2\sigma$, as taken from Table~I.%
%.....................
\footnote{For definiteness, Fig.~2 shows the daytime probability
of $^8$B neutrinos, averaged over their production region in the
Sun.}
%.....................
It can be seen that decoherence effects, to some extent, mimic the
effects of a larger mixing. For instance, the curve with $n=0$ in
the right panel is not much different from the curve at
$\sin^2\theta_{12}=0.371$ (upper $2\sigma$ value) in the left
panel. As a consequence, one expects some degeneracy between the
parameters $\gamma_0$ and $\sin^2\theta_{12}$ when fitting the
data. The degeneracy is only partial however, because decoherence
effects can significantly change both the shape and the slope of
the energy profile within the current $2\sigma$ bounds, as evident
in the right panel. Therefore, future measurements of the
(currently not well constrained) solar neutrino energy spectrum
will provide further important probes of decoherence effects.

The variation of $P_{ee}^\odot$ due to subdominant
decoherence effects [see Eqs.~(\ref{sol_osc}) and (\ref{sol_osc_dec})]
is given, in first approximation, by
%.................................................................
\begin{equation}
\Delta P_{ee}^\odot \simeq -\frac{1}{2}\,\gamma_\odot
R_\odot\cos2\tilde\theta_{12}\cos2\theta_{12}\ ,
\end{equation}
%.................................................................
and changes sign with $\cos2\tilde\theta_{12}$. As the energy
increases, the value of $\cos2\tilde\theta_{12}$ changes from
$\cos \theta_{12}>0$ (low-energy, vacuum-dominated regime) to $-1$
(high-energy, matter-dominated regime), the transition being
located around 2~MeV for the $^8$B neutrino curves shown in
Fig.~2. This fact explains the general increase of $P_{ee}^\odot$
for $E\gtrsim 2$ MeV in the right panel of Fig.~2. More detailed
features depend instead on the energy behavior of the function
$g_n(E)$, which modulates decoherence effects (see the Appendix).
In general, $g_n(E)$ grows rapidly with increasing energy for
$n>0$ (which explains the high-energy upturn of the curves with
$n=+1$ and $n=+2$), while it vanishes with decreasing energy for
all $n\neq -2$
(which explains the low-energy equality of all curves but for $n=-2$).%
%...................................
\footnote{Constraints on the specific case $n=-2$ might thus benefit
of sub-MeV solar neutrino observations in Borexino \protect\cite{Borex}.}
%..................................
 The ``bunching'' of the curves
around $E\sim 10$ MeV in the right panel of Fig.~2 is in part a
data selection effect, since this energy region
is strongly constrained by precise $^8$B neutrino data.
Further spectral $^8$B data will be very useful
to constrain the slope of the energy spectrum and thus also
the sign of the power-law index $n$.

Figure~3 illustrates the partial degeneracy between decoherence
effects and mixing angle, as a shift in the allowed regions for
fixed $\gamma_0 \neq 0$ in three representative cases (from left
to right, $n=-2,\,0,\,+2$). In each panel, the thin dotted curves
enclose the mass-mixing parameter regions allowed at $2\sigma$ by
the standard oscillation fit of solar data (larger region) and by
solar plus KamLAND data (smaller region). The thick solid curves
refer to the same data, but fixing {\em a priori\/} the
decoherence parameter $\gamma_0$ at the $2\sigma$ upper limit
value in Table~I. In all cases in Fig.~3, the curves with
$\gamma_0\neq 0$ are shifted to lower values of the mixing angle,
as compared to pure oscillations; this means that decoherence
effects can be partly traded for a smaller value of the mixing
angle in solar neutrino oscillations. Therefore, should future
solar neutrino data prefer smaller (larger) values of
$\sin^2\theta_{12}$ with respect to KamLAND data, there would be
more (less) room for possible subdominant decoherence effects. As
already remarked, the degeneracy between $\gamma_0$ and
$\sin^2\theta_{12}$ is only partial, and future neutrino
spectroscopy will provide a further handle to break it, should
decoherence effects (if any) be found.

We conclude this section by confronting the bounds in Table~I with
those derivable from the analysis of atmospheric neutrino data
(which, by themselves, exclude pure decoherence, at least in the
$n=-1$ case \cite{SKLE}). In principle, a direct comparison is not
possible, since the $\gamma_0$ parameter introduced here for the
solar $\nu_e\to\nu_e$ channel does not need to be the same as for
the atmospheric $\nu_\mu\to\nu_\mu$ channel. However, if the
$\gamma_0$'s for these two channels are {\em assumed\/} to be
roughly equal in size, then it is easy to realize that
solar+KamLAND neutrinos set stronger (weaker) bounds than
atmospheric neutrinos for $n<0$ ($n>0$), as a consequence of the
different neutrino energy range probed. In fact, due to the
assumed power-law energy dependence of decoherence effects,
negative (positive) values of $n$ are best probed by low-energy
solar (high-energy atmospheric) neutrino experiments. For the
intermediate case $(n=0)$, it turns out that matter effects render
solar neutrinos more sensitive to $\gamma_0$ than atmospheric
neutrinos. Just to make specific numerical examples: for
$n=(-1,\,0,+2)$ one roughly gets the bounds
$\gamma_0\lesssim(0.7\times 10^{-21},\,0.4\times 10^{-22},\,
0.9\times 10^{-27})$ GeV from atmospheric \cite{Lisi1} (plus
accelerator \cite{Lisi2}) neutrino data, to be compared with the
corresponding limits from solar+KamLAND data from Table~I,
$\gamma_0<(0.78\times 10^{-26},\,0.67\times 10^{-24},\,0.47\times
10^{-20})$~GeV. Solar neutrinos clearly win over atmospheric
neutrinos for $n\leq 0$. This comparison must be taken with a
grain of salt, since it can radically change by assuming either
independent $\gamma_0$'s in different oscillation channels, or
functional forms of $\gamma(E)$ different from power laws.

Finally we observe that in the $n=+2$ case, motivated by some
``quantum-gravity'' or ``string-inspired'' models~\cite{string},
the solar+KamLAND limit on $\gamma_0$ is one order of magnitude
lower than the theoretical expectation [$\gamma_0\sim
O(E_0^2/M_{\rm Planck})\sim 10^{-19}$~GeV]. In the case of
atmospheric neutrinos, this bound is even stronger
($\gamma_0<0.9\times 10^{-27}$~GeV). Consequently, these models
appear strongly disfavored, at least in the neutrino sector.

\section{Recovering the case of density fluctuations in the sun}

Decoherence effects in solar neutrino oscillations can be induced
not only by quantum gravity, but also by more ``prosaic''
sources, such as matter density fluctuations---possibly induced by
turbulence in the innermost regions of the Sun. This topic has
been widely investigated in the literature
\cite{Lore94,Nuno96,Bala96,Burg97,Byko00,Burg03,Guzz03,Bala03,Burg04,Bena04},
and quantitative upper limits have already been
set~\cite{Burg03,Guzz03,Bala03,Burg04} by combining solar data and
 first KamLAND results.

It turns out that stochastic density fluctuations lead
(with appropriate redefinition of parameters) to effects
which have the same functional form as those induced by quantum
gravity in the $n=-2$ case. More precisely, let us
consider fluctuations of the solar electron density
$N_e$ around the average value $\langle N_e \rangle$
predicted by the standard solar model,
%%%%%%%%%%%%%%%%%%%%%%%%%%%%%%%%%%%%%%%%%%%%%%%%%%%%%%%%%%%%%%%%%%%
\begin{equation}
N_e (r) = (1 + \beta F (r)) \langle N_e (r) \rangle , \label{dens}
\end{equation}
%%%%%%%%%%%%%%%%%%%%%%%%%%%%%%%%%%%%%%%%%%%%%%%%%%%%%%%%%%%%%%%%%%%
where $F (r)$ is a random variable describing fluctuations at
a given radius $r$, and $\beta$ represents their fractional
amplitude around the average. It is customary to assume a
delta-correlated (white) noise,
%%%%%%%%%%%%%%%%%%%%%%%%%%%%%%%%%%%%%%%%%%%%%%%%%%%%%%%%%%%%%%%%%%%
\begin{equation}
\langle F (r_1) F (r_2) \rangle =  2\tau \delta (r_1-r_2),
\label{eq:delta}
\end{equation}
%%%%%%%%%%%%%%%%%%%%%%%%%%%%%%%%%%%%%%%%%%%%%%%%%%%%%%%%%%%%%%%%%%%
where $\tau$ is the correlation length of fluctuations along the
($\sim$ radial) neutrino direction.

As shown in~\cite{Burg97} through a perturbative method (which
inspired our approach to decoherence in the Appendix), the effect
of delta-correlated noise on adiabatic neutrino flavor transitions
can be embedded through an exponential damping factor as in
Eq.~(\ref{sol_osc_dec}). The functional form turns out to be the
same as for the $n=-2$ case, provided that one makes---in our
notation---the replacement
%................................
\begin{equation}
\gamma_0 \to \beta^2\tau \left(\frac{\delta m^2}{2E_0}\right)^2\ .
\end{equation}
%..................................
By using the bound $\gamma_0<0.81\times 10^{-28}$~GeV (Table~I,
case $n=-2$) and the best-fit value $\delta m^2=7.92\times
10^{-5}$~eV$^2$, one gets the following upper limit
%%%%%%%%%%%%%%%%%%%%%%%%%%%%%%%%%%%%%%%%%%%%%%%%%%%%%%%%%%%%%%%%%%%%
\begin{equation}
\beta^2 \tau < 1.02\times 10^{-2}\, \mathrm{km} ~(2\sigma)
\,,\label{eq:beta2tau}
\end{equation}
%%%%%%%%%%%%%%%%%%%%%%%%%%%%%%%%%%%%%%%%%%%%%%%%%%%%%%%%%%%%%%%%%%%%
on the parameter combination $\beta^2 \tau$ which is relevant \cite{Burg97,Bala03} for density
fluctuation effects on neutrino propagation.

As stressed in~\cite{Bala03},
care must be taken in extracting an upper limit on the fractional
amplitude $\beta$ for fixed correlation length $\tau$. Indeed, the
delta-correlated noise is an acceptable approximation only if the
correlation length $\tau$ is much smaller than the oscillation
wavelength in matter---a condition that becomes critical for low
neutrino energies. In particular, assuming a reference value $\tau =
10$~km as in~\cite{Bala03}, such condition is violated
for energies $E \lesssim 1$~MeV. Following  \cite{Bala03}, we have thus
excluded low-energy, radiochemical solar neutrino data
form the solar+KamLAND data fit, and obtained a slightly weaker
(but more reliable) upper bound from the analysis of $^8$B neutrino data
only,
%%%%%%%%%%%%%%%%%%%%%%%%%%%%%%%%%%%%%%%%%%%%%%%%%%%%%%%%%%%%%%%%%%%%
\begin{equation}
\beta^2 \tau < 1.16\times 10^{-2}\, \mathrm{km} ~(2\sigma)
\,,\label{eq:beta2tau2}
\end{equation}
%%%%%%%%%%%%%%%%%%%%%%%%%%%%%%%%%%%%%%%%%%%%%%%%%%%%%%%%%%%%%%%%%%%%
which,  for $\tau=10$~km, translates into an upper limit on the
fractional fluctuation
amplitude,
%%%%%%%%%%%%%%%%%%%%%%%%%%%%%%%%%%%%%%%%%%%%%%%%%%%%%%%%%%%%%%%%%%%%
\begin{equation}
\beta < 3.4\%
~(2\sigma) \,.\label{eq:betaLimit}
\end{equation}
%%%%%%%%%%%%%%%%%%%%%%%%%%%%%%%%%%%%%%%%%%%%%%%%%%%%%%%%%%%%%%%%%%%%
This limit improves the previous one derived in \cite{Bala03} ($\beta <
6.3\%$ at $2\sigma$) by a factor of $\sim 2$, essentially as a result
of the inclusion of the most recent solar
and KamLAND neutrino data appeared after
\cite{Bala03}.%
\footnote{We have verified that, by adopting the same (older) data set and
standard solar model as used in \protect\cite{Bala03}, we recover
the same $2\sigma$ upper limit, $\beta\lesssim 6\%$.}

\section{Summary and conclusions}

In this paper, we have investigated
hypothetical decoherence effects (e.g., induced by quantum gravity)
in the $\nu_e \to \nu_e$ oscillation channel explored by the solar
and KamLAND experiments. In both kinds of of experiments, decoherence
effects can be embedded through exponential damping
factors, proportional to a common parameter $\gamma_0$,
which modulate the energy-dependent
part of the $\nu_e$ survival probability. By assuming that
the (unknown) functional form of decoherence effects is a power-law in
energy $(E^n)$, we have studied
the phenomenological
constraints on the main decoherence parameter
($\gamma_0$) for $n=0,\,\pm 1,\,\pm 2$.
It turns out that both solar and KamLAND data do not provide indications in favor of decoherence effects and prefer the standard oscillation case ($\gamma_0=0$)
for any index $n$. By combining the two data sets, $n$-dependent
upper bounds (dominated by solar neutrino data)
have been derived on $\gamma_0$, as reported in Table~I and shown in Fig.~1.
In the ``Lorentz-invariant'' case $n=-1$, we obtain
the upper limit $\gamma_0<0.78\times 10^{-26}$ GeV at $95\%$ C.L.
For $n=-2$, the results can also be interpreted as limits on the amplitude
of possible (delta-correlated) density fluctuations in the Sun, which
we improve by a factor of two [Eqs.~(\ref{eq:beta2tau2}) and (\ref{eq:betaLimit})] with respect
to previous bounds.

Further progress might come from a better determination of the
energy profile of solar neutrino flavor transitions as well as
from more precise measurements of $\sin^2\theta_{12}$ (which is
partly degenerate with $\gamma_0$), attainable with KamLAND and
future long-baseline reactor neutrino experiments \cite{Petc}.

\section*{APPENDIX: DECOHERENCE AND MATTER EFFECTS IN SOLAR NEUTRINOS}

In this section we discuss a perturbative calculation of
decoherence effects for solar neutrinos, where matter effects are
known to be relevant. The approach, inspired by the
work~\cite{Burg97}, draws on the formalism and the notation
introduced in~\cite{Lisi1} for the case of decoherence in the
$\nu_\mu\to\nu_\mu$ channel, here adapted to the $\nu_e\to\nu_e$
channel. In the following, the the notation is made more compact
by setting $\theta=\theta_{12}$, $c_{2\theta}=\cos 2\theta_{12}$,
$s_{2\theta}= \sin 2\theta_{12}$, $P_{ee}=P_{ee}^\odot$ etc.

Decoherence effects in the
flavor evolution of the $(\nu_e,\nu_a)$ system (where $a=\mu,\tau$)
along the space coordinate $r(\simeq t$)%
%%%%%%%%%%%%%%%%%%%%%%%%%%%%%%%%%%%%%%%%%%%%%%%%%
\footnote{Note that $r$ does not necessarily coincide with the
radial coordinate, due to the extended neutrino production region
(which is taken into account in our analysis.)}
%%%%%%%%%%%%%%%%%%%%%%%%%%%%%%%%%%%%%%%%%%%%%%%%%
can be described in terms of the neutrino density matrix, obeying
a modified master equation of the form~\cite{Lind76}
\begin{equation}
\frac{d\rho}{dr} = -i[H_v+H_m(r),\rho]-\gamma [D,[D,\rho]]\ ,
\label{eq:lindblad}\end{equation}
where $H_v$ ($H_m$) is the
``vacuum'' (matter) Hamiltonian, and the operator $D$ embeds
decoherence effects with amplitude $\gamma$, parameterized as
in Eq.~(\ref{eq:gamma}): $\gamma=\gamma_0(E/E_0)^n$ with $E_0=1$~GeV.
While unitarity is preserved (i.e., $\mathrm{Tr}\rho(r)=1$),
coherence is lost in the propagation
($\frac{d}{dr}\mathrm{Tr}\rho^2\leq 0$). Equation
~(\ref{eq:lindblad}) satisfies the conditions of
complete positivity~\cite{Gori76}
and non-decreasing entropy in the $\nu$ system evolution~\cite{Bena88}.

In the flavor basis the standard oscillation terms read
%%%%%%%%%%%%%%%%%%%%%%%%%%%%%%%%%%%%%%%%%%%%%%%%%%%%%%%%%%%%%%%%%%%%%
\begin{eqnarray}
H_v &=& -\frac{k}{2}U_\theta \sigma_3 U^\dag_\theta =
\frac{k}{2}\left[\begin{array}{cc}
-c_{2\theta} & s_{2\theta}\\
 s_{2\theta} & c_{2\theta}
\end{array}\right]\ , \\
H_m &=& -\frac{V(r)}{2} \sigma_3\ , \label{eq:hamiltonian}
\end{eqnarray}
%%%%%%%%%%%%%%%%%%%%%%%%%%%%%%%%%%%%%%%%%%%%%%%%%%%%%%%%%%%%%%%%%%%%%
where $\sigma_3$ is the third Pauli matrix, $k=\delta m^2/2E$ is the vacuum wavenumber, and
$V(r)=\sqrt{2}G_F N_e(r)$ is the interaction potential in matter.

As in~\cite{Lisi1}, we assume
energy conservation for evolution in vacuum,
i.e., $\mathrm{Tr}[H_v\rho(r)]=$~constant.
This condition is satisfied
if $[H_v,D]=0$~\cite{Banks84,Jliu}, namely, in a two-dimensional system, if
$D\propto H_v$. We can thus take
%%%%%%%%%%%%%%%%%%%%%%%%%%%%%%%%%%%%%%%%%%%%%%%%%%%%%%%%%%%%%%%%%%%%%
\begin{equation}
D_\theta=\frac{1}{2}\left[\begin{array}{cc}
-c_{2\theta} & s_{2\theta}\\
 s_{2\theta} & c_{2\theta}
\end{array}\right]\ ,\label{eq:decoherence}
\end{equation}
%%%%%%%%%%%%%%%%%%%%%%%%%%%%%%%%%%%%%%%%%%%%%%%%%%%%%%%%%%%%%%%%%%%%%
without loss of generality, since any overall
factor can be absorbed in $\gamma_0$. We also make the plausible
assumption that Eq.~(\ref{eq:decoherence}) is not altered for
evolution in matter, since decoherence
induced by quantum gravity is unrelated to electroweak matter effects.

As usual, Eq.~(\ref{eq:lindblad}) can be written in terms of a
``polarization'' vector $\mathbf{P}$ with components $P_i=\frac{1}{2}\mathrm{Tr}[\rho\sigma_i]$
(Bloch equation):
%%%%%%%%%%%%%%%%%%%%%%%%%%%%%%%%%%%%%%%%%%%%%%%%%%%%%%%%%%%%%%%%%%%%%
\begin{eqnarray}
\frac{d\mathbf{P}}{dr} &=&
\left[k\mathbf{n}+V(r)\mathbf{e}_3\right]\times
\mathbf{P}-\gamma\mathbf{P}_\perp \nonumber \\
&=& \mathcal{H}(r)\mathbf{P}-\gamma
\mathcal{D}_\theta\mathbf{P}
\,\label{eq:bloch}
\end{eqnarray}
%%%%%%%%%%%%%%%%%%%%%%%%%%%%%%%%%%%%%%%%%%%%%%%%%%%%%%%%%%%%%%%%%%%%%
where $\mathbf{n}=[s_{2\theta},0,-c_{2\theta}]^T$,
 $\mathbf{P}_\perp=\mathbf{P}-(\mathbf{P}\cdot\mathbf{n})\mathbf{n}$, and:
%%%%%%%%%%%%%%%%%%%%%%%%%%%%%%%%%%%%%%%%%%%%%%%%%%%%%%%%%%%%%%%%%%%%%
\begin{equation}
\mathcal{H}(r) =
\left[\begin{array}{ccc}
0                 & -V(r)+kc_{2\theta} & 0\\
V(r)-kc_{2\theta} & 0                  & -ks_{2\theta}\\
0                 & ks_{2\theta}       & 0
\end{array}\right]\ ,\label{eq:H}
\end{equation}
\begin{equation}
\mathcal{D}_\theta =
\left[\begin{array}{ccc}
c^2_{2\theta}          & 0 & c_{2\theta}s_{2\theta}\\
0                      & 1 & 0\\
c_{2\theta}s_{2\theta} & 0 & s^2_{2\theta}
\end{array}\right]\ .\label{eq:D}
\end{equation}
%%%%%%%%%%%%%%%%%%%%%%%%%%%%%%%%%%%%%%%%%%%%%%%%%%%%%%%%%%%%%%%%%%%%%
Note that for $V=0$ (vacuum propagation), the solution of the
above Bloch equation leads to the survival probability in
Eq.~(\ref{K_osc_dec}), (see~\cite{Lisi1} for details.)

The matrix $\mathcal{H}$ has eigenvalues $\lambda_0=0$ and
$\lambda_\pm=\pm \tilde k$, corresponding to the eigenvectors:
%%%%%%%%%%%%%%%%%%%%%%%%%%%%%%%%%%%%%%%%%%%%%%%%%%%%%%%%%%%%%%%%%%%%%
\begin{equation}
\begin{array}{ccc}
\mathbf{u}_0=\left[\begin{array}{c}
s_{2\tilde \theta}\\
0\\
-c_{2\tilde\theta}
\end{array}\right]\ , & &
\mathbf{u}_\pm=\frac{1}{\sqrt{2}}\left[\begin{array}{c}
c_{2\tilde\theta}\\
\pm i\\
s_{2\tilde\theta}
\end{array}\right]
\end{array}\ .\label{eq:eigenvectors}
\end{equation}
%%%%%%%%%%%%%%%%%%%%%%%%%%%%%%%%%%%%%%%%%%%%%%%%%%%%%%%%%%%%%%%%%%%%%
In the above equations, a ``tilde'' marks effective parameters in
matter: $\tilde k$ is the oscillation wavenumber in matter, defined
through
 $\tilde k/k=[{1-2Vc_{2\theta}/k+(V/k)^2}]^{1/2}$, while $\tilde\theta$ is the mixing
angle in matter, defined through
$s_{2\tilde\theta}=ks_{2\theta}/\tilde k$ and
$c_{2\tilde \theta}=(kc_{2\theta}-V)/\tilde k$. The matrix $\mathcal{H}$
is diagonalized through the matrix
$\mathcal{R}(\tilde \theta)=[\mathbf{u}_0,\mathbf{u}_+,\mathbf{u}_-]$: $\mathcal{R}^\dag(\tilde \theta)
\cdot\mathcal{H}\cdot\mathcal{R}(\tilde \theta)=\mathrm{diag}[0,+i\tilde k,
-i\tilde k]$.

In the absence of decoherence effects, the adiabatic solution of
Eq.~(\ref{eq:bloch}) appropriate for current solar neutrino
phenomenology is
%%%%%%%%%%%%%%%%%%%%%%%%%%%%%%%%%%%%%%%%%%%%%%%%%%%%%%%%%%%%%%%%%%%%%
\begin{equation}
\mathbf{P}(R_\odot)=\mathcal{R}(\theta)\cdot\mathrm{diag}
\left[1,\,e^{+i\int_{r_0}^{R_\odot} dr\, \tilde k(r)},\,
e^{-i\int_{r_0}^{R_\odot} dr\,
\tilde k(r)}\right]\cdot
\mathcal{R}^\dag(\tilde \theta_0)\mathbf{P}(r_0)\, \label{eq:adiabatic}
\end{equation}
where $\mathbf{P}(r_0)=\,^T[0,0,1]$ for an initial $\nu_e$ state, $\tilde\theta_0$ is the mixing angle in matter at the production point $r_0$,
and $\tilde\theta=\theta$ is taken at $r=R_\odot$.
After averaging on the fast oscillating terms
(a ``standard'' decoherence effect), one recovers the
usual adiabatic formula [Eq.~(\ref{sol_osc})]
for the survival probability $P_{ee}$
\begin{equation}
P_{ee}=\mathrm{Tr}\left[\rho|\nu_e\rangle\langle\nu_e|\right]=\frac{1+P_3(R_\odot)}{2}=\frac{1+c_{2\theta}c_{2\tilde\theta_0}}{2}\ .\label{eq:parke3}
\end{equation}
%%%%%%%%%%%%%%%%%%%%%%%%%%%%%%%%%%%%%%%%%%%%%%%%%%%%%%%%%%%%%%%%%%%%%

Let us now treat the term
$-\gamma \mathcal{D}_\theta\mathbf{P}$ in Eq.~(\ref{eq:bloch}) as
a perturbation~\cite{Burg97}. The corrections to the
eigenvectors lead to variations of $P_{ee}$ of
$O(\gamma/k)\lesssim 10^{-3}$ (for the range of $\gamma/k$ allowed
a posteriori by the fit to solar neutrino data) and can
be neglected. The
corrections to the two eigenvalues $\lambda_\pm$ can also be neglected,
since they would only lead
to a further damping of the fast oscillating terms, which are already
averaged out.%
%----------------------------------------------
\footnote{Similarly, non-standard decoherence effects in the path
from the Sun surface to the Earth are irrelevant, since they would
simply damp the (already averaged out) fast oscillations. The only
relevant effects occur within the Sun, through the modification of
the standard evolution in matter.}
%----------------------------------------------

The first-order
correction to the eigenvalue $\lambda_0$ (whose unperturbed value is zero)
is the only relevant one,
%%%%%%%%%%%%%%%%%%%%%%%%%%%%%%%%%%%%%%%%%%%%%%%%%%%%%%%%%%%%%%%%%%%%%
\begin{equation}
\delta \lambda_0=-\gamma \mathbf{u}_0^\dag\mathcal{D}_\theta\mathbf{u}_0=-\gamma\sin^2 2(\tilde \theta-\theta)\ .\label{eq:lambdanew}
\end{equation}
%%%%%%%%%%%%%%%%%%%%%%%%%%%%%%%%%%%%%%%%%%%%%%%%%%%%%%%%%%%%%%%%%%%%%
and leads to the following correction to Eq.~(\ref{eq:parke3}):
%%%%%%%%%%%%%%%%%%%%%%%%%%%%%%%%%%%%%%%%%%%%%%%%%%%%%%%%%%%%%%%%%%%%%
\begin{equation}
P_{ee}=\frac{1+e^{-\Gamma}c_{2\theta}c_{2\tilde\theta_0}}{2}\
,\label{eq:parkenew}
\end{equation}
%%%%%%%%%%%%%%%%%%%%%%%%%%%%%%%%%%%%%%%%%%%%%%%%%%%%%%%%%%%%%%%%%%%%%
where
%%%%%%%%%%%%%%%%%%%%%%%%%%%%%%%%%%%%%%%%%%%%%%%%%%%%%%%%%%%%%%%%%%%%%
\begin{equation}
\Gamma= \gamma \int_{r_0}^{R_\odot}
\left[\frac{V(r)s_{2\tilde\theta}(r)}{k}\right]^2 dr\ .
\label{eq:gamma2}
\end{equation}
%%%%%%%%%%%%%%%%%%%%%%%%%%%%%%%%%%%%%%%%%%%%%%%%%%%%%%%%%%%%%%%%%%%%%
Equation~(\ref{sol_osc_dec}) is then recovered by setting
$\gamma_\odot=\gamma_0\,g_n(E)$ and by defining
the dimensionless function $g_n(E)$ as
%..........................
\begin{equation}
g_n(E)=\left(\frac{E}{E_0}\right)^n
\int_{r_0}^{R_\odot} \left[\frac{V(r)s_{2\tilde\theta}(r)}{k}\right]^2
\frac{dr}{R_\odot}\ .
\end{equation}
%..............................

The function $g_n(E)$ depends mostly on the neutrino energy $E$
and, to some extent,
on the parameters $r_0$, $\delta m^2$, and $\sin^2 \theta_{12}$.
Figure~4 shows this
function as calculated for $r_0=0$, $\delta m^2=7.92\times 10^{-5}$ eV$^2$,
and $\sin^2\theta_{12}=0.314$. For
$E\to 0$ the function $g_n(E)$ (and the associated decoherence effect)
vanishes, except for the case $n=-2$, where
the factors $E^{-2}$ and $k^{-2}$
cancel out and provide a finite limit $g_{-2}(0)\neq 0$.

We have tested the analytical Eq.~(\ref{eq:parkenew}) against the
results of a numerical integration of the Bloch equation, for many
representative points in the parameter space relevant for solar
$\nu$ phenomenology, and we find very good agreement ($\delta
P_{ee}< 10^{-4}$) for all values of $n\neq -2$. Only in the case
$n=-2$, the comparison of analytical and numerical results is
slightly worse (but still very good, $\delta P_{ee}< 10^{-3}$) at
the lowest detectable energies $(\sim 0.1$~MeV), due to the
breakdown of perturbation theory for $E\to 0$. For practical
purposes, however, the modified adiabatic Eq.~(\ref{eq:parkenew})
accurately replaces the results of numerical solutions of the
Bloch equation for solar neutrinos.

\acknowledgments

The work of G.L.\ Fogli, E.\ Lisi, A.\ Marrone, and D.\ Montanino
is supported by the Italian MUR and INFN through the
``Astroparticle Physics'' research project. The work of A.\
Palazzo is supported by INFN.

\newpage

%%%%%%%%%%%%%%%%%%%%%%%%%%%%%%%%%%%%%%%%%%%%%%%%%%%%%%%%%%%%%%%%%%%%%%%%%
\begin{figure}
\includegraphics[scale=0.90, bb= 100 100 510 720]{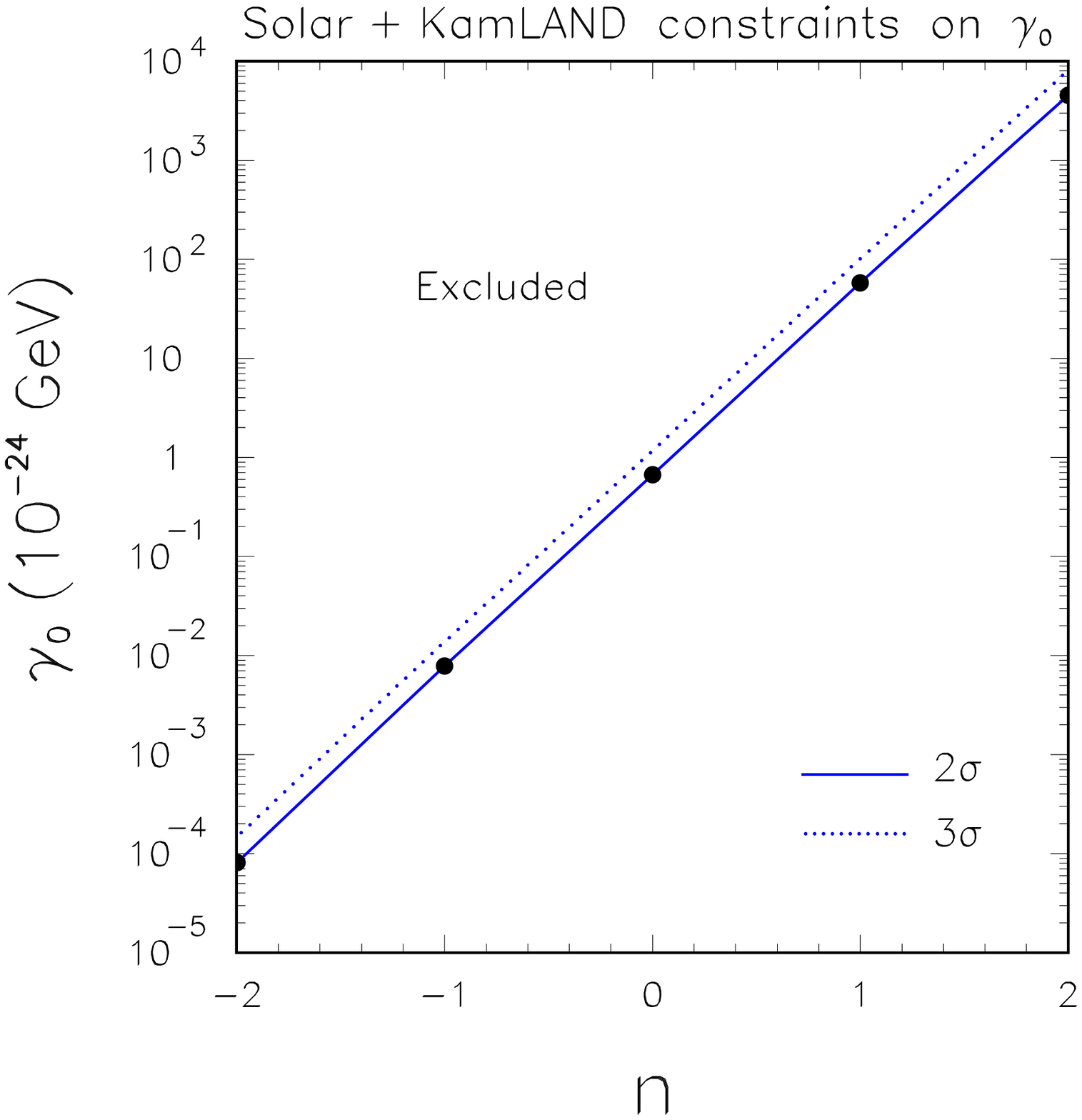}
\vspace*{-1.0cm} \caption{ \label{Fig_1} Upper bounds on the
decoherence parameter $\gamma_0$ as a function of the power-law
index $n$, as obtained from a combined analysis of solar and
KamLAND data. The solid and dotted curves refer to $2\sigma$ and
$3\sigma$ confidence level, respectively.}
\end{figure}
%%%%%%%%%%%%%%%%%%%%%%%%%%%%%%%%%%%%%%%%%%%%%%%%%%%%%%%%%%%%%%%%%%%%%%%%%
%%%%%%%%%%%%%%%%%%%%%%%%%%%%%%%%%%%%%%%%%%%%%%%%%%%%%%%%%%%%%%%%%%%%%%%%%
\begin{figure}
\includegraphics[scale=0.90, bb= 100 100 510 720]{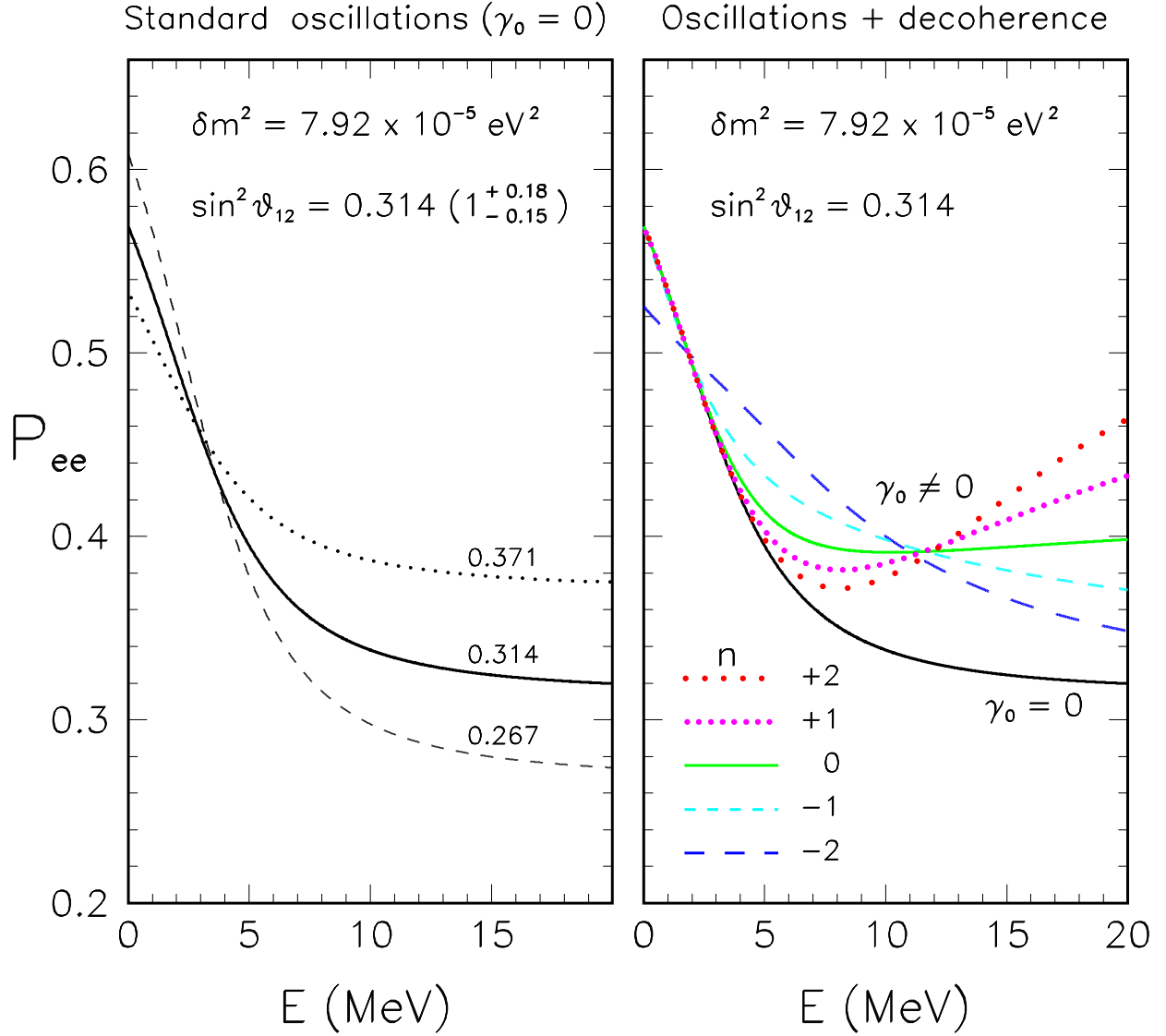}
\vspace*{-1.0cm} \caption{ \label{Fig_2} Energy profile of the
(daytime) survival probability of $^8$B neutrinos, averaged over
their production region in the sun: Comparison of the effects
produced by variations of $\sin^2\theta_{12}$ for $\gamma_0 = 0$
(left panel) and by $\gamma_0\neq 0$ at fixed $\sin^2\theta_{12}$
(right panel). In the left panel, $\sin^2\theta_{12}$ is varied
within its $\pm2\sigma$ limits. In the right panel, for each index
$n =[-2,-1,0, +1, +2]$ the value of $\gamma_0$ is taken equal to
the corresponding $2\sigma$ upper limit (reported in Table~I),
which in units of $10^{-24}$~GeV corresponds respectively to:
$0.81 \times 10^{-4}$ $(n= -2)$, $0.78 \times 10^{-2}$ $(n= -1)$,
$0.67$ $(n= 0)$, $0.58 \times 10^{2}$ $(n= +1)$, $0.47 \times
10^{4}$ $(n= +2)$. The curve corresponding to standard
oscillations ($\gamma_0 = 0$) is also shown in the right panel as
a guide to the eye. In all cases, $\delta m^2$ is fixed at its
best-fit value.}
\end{figure}
%%%%%%%%%%%%%%%%%%%%%%%%%%%%%%%%%%%%%%%%%%%%%%%%%%%%%%%%%%%%%%%%%%%%%%%%%
%%%%%%%%%%%%%%%%%%%%%%%%%%%%%%%%%%%%%%%%%%%%%%%%%%%%%%%%%%%%%%%%%%%%%%%%%
\begin{figure}
\includegraphics[scale=0.90, bb= 100 100 510 720]{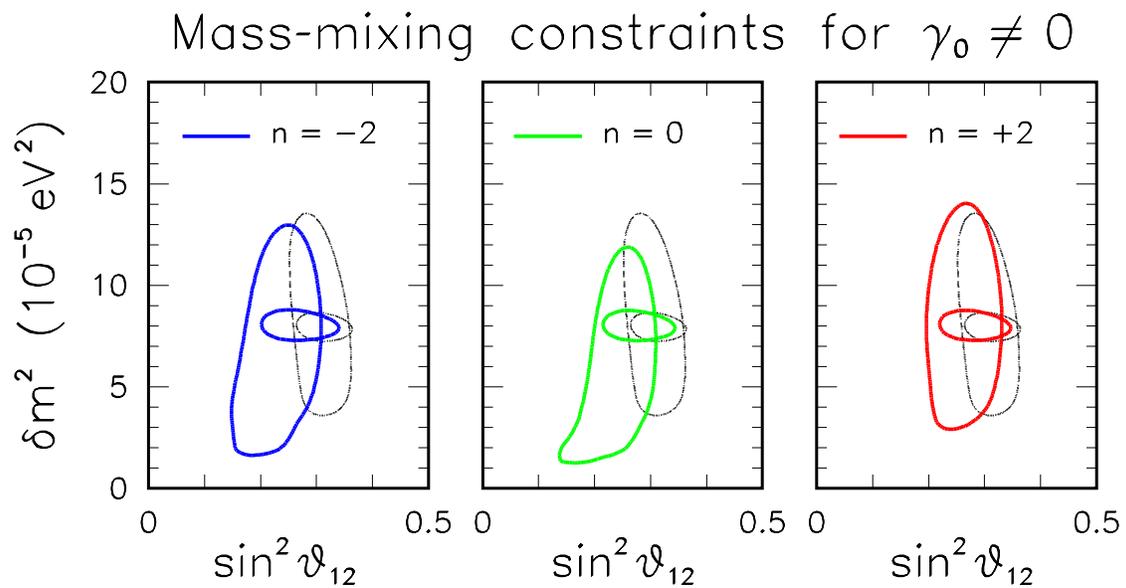}
\vspace*{-5.0cm} \caption{ \label{Fig_3} Constraints on the
mass-mixing parameters for $\gamma_0=0$ (thin dotted curves) and
for $\gamma_0$ fixed at its $2\sigma$ upper limit in Table~I
(thick solid curves). The three panels refer, from left to right,
to the three cases $n=-2$, 0, and $+2$. The smaller (larger)
allowed regions refer to the solar+KamLAND (solar only) data
analysis.}
\end{figure}
%%%%%%%%%%%%%%%%%%%%%%%%%%%%%%%%%%%%%%%%%%%%%%%%%%%%%%%%%%%%%%%%%%%%%%%%%
%%%%%%%%%%%%%%%%%%%%%%%%%%%%%%%%%%%%%%%%%%%%%%%%%%%%%%%%%%%%%%%%%%%%%%%%%
\begin{figure}
\includegraphics[scale=0.90, bb= 100 100 510 720]{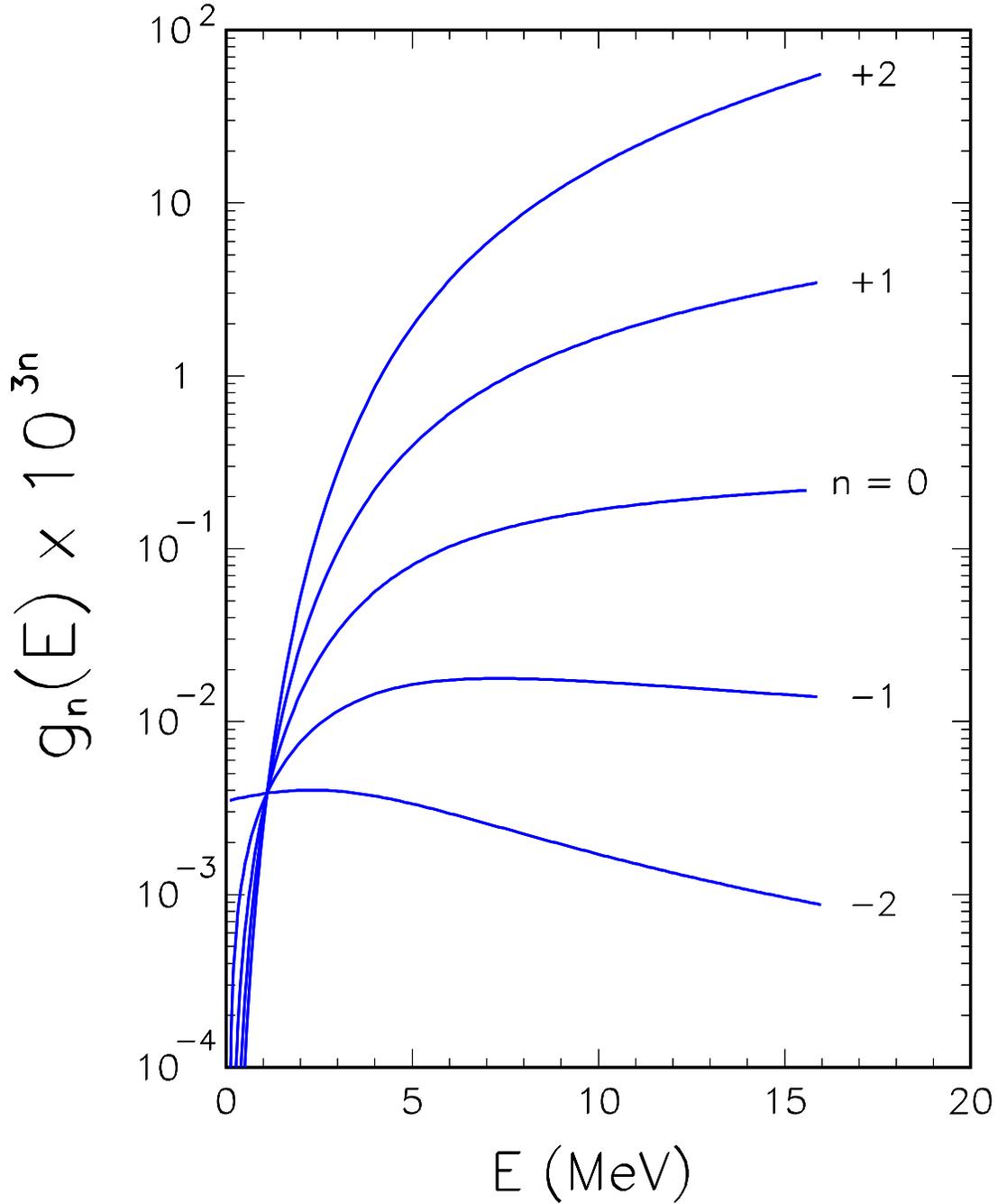}
\vspace*{-.0cm} \caption{ \label{Fig_4} Energy profile of the auxiliary function
$g_n(E)$, which modulates the exponent of the damping factor induced
by decoherence in solar neutrino oscillations. The function is multiplied
by $10^{3n}$ for a better graphical view. The shown function refers
to a neutrino produced at the Sun center and to best-fit oscillation
parameters; altering this choice would induce minor variations.
See the text for details.}
\end{figure}
%%%%%%%%%%%%%%%%%%%%%%%%%%%%%%%%%%%%%%%%%%%%%%%%%%%%%%%%%%%%%%%%%%%%%%%%%


\begin{thebibliography}{99}



\bibitem{Haw75}     S.W.\ Hawking,
                    Commun.\ Math.\ Phys.\ {\bf 43}, 199 (1975);
                    {\em ibidem\/} {\bf 87}, 395 (1982);
                    Phys.\ Rev.\ D {\bf 14}, 2460 (1976).


\bibitem{Foam}      S.B.\ Giddings and A.\ Strominger,
                    Nucl.\ Phys.\ B {\bf 307}, 854 (1988);
                    W.H.\ Zurek, Physics Today {\bf 44}, No.~10, p.~36 (1991);
                    G.\ Amelino-Camelia, J.\ Ellis, N.E.\ Mavromatos, and
                    D.V.\ Nanopoulos,
                    Int.\ J.\ Mod.\ Phys.\ A {\bf 12}, 607 (1997);
                    L.J.\ Garay, Int.\ J.\ Mod.\ Phys.\ A {\bf 14}, 4079 (1999).


\bibitem{El84}      J.\ Ellis, J.S.\ Hagelin, D.V.\ Nanopoulos, and M.\ Srednicki,
                    Nucl.\ Phys.\ B {\bf 241}, 381 (1984).


\bibitem{Banks84}   T.\ Banks, L.\ Susskind, and M.E.\ Peskin,
                    Nucl.\ Phys.\ B {\bf 244}, 125 (1984).

\bibitem{kaon}      P.\ Huet and M.E.\ Peskin,
                    Nucl.\ Phys.\ B {\bf 488}, 335 (1997);
                    J.\ Ellis, J.L\ Lopez, N.E.\ Mavromatos, and D.V.\ Nanopoulos,
                    Phys.\ Rev. D {\bf 53}, 3846 (1996);
                    F.\ Benatti and R.\ Floreanini,
                    Phys.\ Lett.\ B {\bf 389}, 100 (1996);
                    {\em ibidem\/} {\bf 401}, 337 (1997).

\bibitem{Bern06}    J.\ Bernabeu, N.E.\ Mavromatos, and S.\ Sarkar,
                    Phys.\ Rev.\ D {\bf 74}, 045014 (2006).

\bibitem{neut}      F.\ Benatti and R.\ Floreanini,
                    Phys.\ Lett.\ B {\bf 451}, 422 (1999).


\bibitem{sol1}      Y.\ Liu, L.\ Hu, and M.L.\ Ge,
                    Phys.\ Rev.\ D {\bf 56}, 6648 (1997).

\bibitem{sol2}      Y.\ Liu, J.L.\ Chen, and M.L.\ Ge,
                    J.\ Phys.\ G {\bf 24}, 2289 (1998);
                    C.P.\ Sun and D.L.\ Zhou, hep-ph/9808334.

\bibitem{atm1}      C.H.\ Chang, W.S. Dai, X.Q.\ Li, Y.\ Liu,  F.C.\ Ma,
                    and Z.J.\ Tao,
                    Phys.\ Rev.\ D {\bf 60}, 033006 (1999).


\bibitem{Lisi1}     E.\ Lisi, A.\ Marrone, and D.\ Montanino,
                    Phys.\ Rev.\ Lett.\ {\bf 85}, 1166 (2000).

\bibitem{Lisi2}     G.\ L.\ Fogli, E.\ Lisi, A.\ Marrone, and D.\ Montanino,
                    Phys.\ Rev.\ D {\bf 67}, 093006 (2003).


\bibitem{Gago01}    A.\ M.\ Gago, E.M.\ Santos, W.J.C.\ Teves, and R. Zukanovich Funchal,
                    Phys.\ Rev.\ D {\bf 63}, 073001 (2001).

\bibitem{Blenn}     M.~Blennow, T.~Ohlsson, and W.~Winter,
                    JHEP {\bf 0506}, 049 (2005);
                    Eur.\ Phys.\ J.\  C {\bf 49}, 1023 (2007).


\bibitem{Atm_evid}  Super-Kamiokande Collaboration, Y.\ Fukuda {\em et al.},
                    Phys.\ Rev.\ Lett. {\bf 81}, 1562 (1998).


\bibitem{Ahlu01}    D.V.\ Ahluwalia,
                    Mod.\ Phys.\ Lett.\ A {\bf 16}, 917 (2001).


\bibitem{Hoop1_04}  D.\ Hooper, D.\ Morgan, and E.\ Winstanley,
                    Phys.\ Lett.\ B {\bf 609}, 206 (2005).

\bibitem{Hoop1_05}  D.\ Hooper, D.\ Morgan, and E.\ Winstanley,
                    Phys.\ Rev.\ D {\bf 72}, 065009 (2005).

\bibitem{Hoop2_05}  L.A.\ Anchordoqui, H.\ Goldberg, M.C.\ Gonzalez-Garcia, F.\ Halzen, D.\ Hooper, S.\ Sarkar and T.J.\ Weiler,
                    Phys.\ Rev.\ D {\bf 72}, 065019 (2005).

\bibitem{Morg_06}   D.\ Morgan, E.\ Winstanley, J.\ Brunner, and L.F.\ Thompson,
                    Astropart.\ Phys.\ {\bf 25}, 311 (2006).


\bibitem{Bena00}    F.\ Benatti and R.\ Floreanini, JHEP 0002, 032 (2000).

\bibitem{Bena01}    F.\ Benatti and R.\ Floreanini, Phys.\ Rev.\ D {\bf 64}, 085015 (2001).


\bibitem{Bare1_05}  G.\ Barenboim and N.E.\ Mavromatos, JHEP {\bf 0501}, 034 (2005).

\bibitem{Bare2_05}  G.\ Barenboim, N.E.\ Mavromatos, S.\ Sarkar, and A.\ Waldron-Lauda,
                    Nucl.\ Phys.\ B {\bf 758}, 90 (2006).

\bibitem{Bare3_05}  G.\ Barenboim and N.E.\ Mavromatos,
                    Phys.\ Rev.\ D {\bf 70}, 093015 (2004).

\bibitem{Mavro_05}  N.E.\ Mavromatos and S.\ Sarkar,
                    Phys.\ Rev.\ D {\bf 72}, 065016 (2005).

\bibitem{Mavro_06}  S.\ Sarkar, hep-ph/0610010.



\bibitem{Pont}      Z.~Maki, M.~Nakagawa, and S.~Sakata,
                    Prog.\ Theor.\ Phys.\ {\bf 28}, 870 (1962);
                    B.~Pontecorvo, Zh.\ Eksp.\ Teor.\ Fiz.\ {\bf
                    53}, 1717 (1967) [Sov.\ Phys.\ JETP {\bf 26}, 984 (1968)].

\bibitem{Matt}      L.~Wolfenstein,
                    Phys.\ Rev.\ D {\bf 17}, 2369 (1978);
                    S.P.~Mikheev and A.Yu.\ Smirnov,
                    Yad.\ Fiz.\ {\bf 42}, 1441 (1985)
                    [Sov.\ J.\ Nucl.\ Phys.\ {\bf 42}, 913 (1985)].

\bibitem{Adia}      L.\ Wolfenstein,
                    in {\it Neutrino~'78}, 8th International
                    Conference on Neutrino Physics and Astrophysics
                    (Purdue U., West Lafayette, Indiana, 1978), ed.\
                    by E.C.\ Fowler (Purdue U.\ Press, 1978), p.~C3.


\bibitem{CHL}       Homestake Collaboration,
                    B.T.~Cleveland, T.~Daily, R.~Davis Jr., J.R.~Distel,
                    K.~Lande, C.K.~Lee, P.S.~Wildenhain, and
                    J.~Ullman, Astrophys.\ J.\  {\bf 496}, 505 (1998).

\bibitem{Kam}       Kamiokande Collaboration, Y.~Fukuda {\it et al.},
                    Phys.\ Rev.\ Lett.\  {\bf 77}, 1683 (1996).

\bibitem{SAGE}      SAGE Collaboration,
                    J.N.~Abdurashitov {\it et al.},
                    J.\ Exp.\ Theor.\ Phys.\  {\bf 95}, 181 (2002)
                    [Zh.\ Eksp.\ Teor.\ Fiz.\  {\bf 95}, 211 (2002)].

\bibitem{GALL}      GALLEX Collaboration, W.~Hampel {\it et al.},
                    Phys.\ Lett.\ B {\bf 447}, 127 (1999).

\bibitem{GGNO}      Gallium Neutrino Observatory (GNO) Collaboration,
                    M.\ Altmann {\em et al.},
                    Phys.\ Lett.\ B {\bf 616}, 174 (2005).

\bibitem{Gavr}      V.\ Gavrin, talk at {\em Neutrino 2006},
                    XXII International Conference on Neutrino Physics and
                    Astrophysics (Santa Fe, New Mexico, USA, 2006).
                    Website: neutrinosantafe06.com


\bibitem{SKso}      SK Collaboration, S.\ Fukuda {\it et al.},
                    Phys.\ Rev.\ Lett.\  {\bf 86}, 5651 (2001);
                    Phys.\ Rev.\ Lett.\  {\bf 86}, 5656 (2001);
                    Phys.\ Lett.\ B {\bf 539}, 179 (2002).

\bibitem{SK04}      SK Collaboration,
                    M.B.\ Smy {\it et al.},
                    Phys.\ Rev.\ D {\bf 69}, 011104 (2004).

\bibitem{SNO1}      SNO Collaboration,
                    Q.R.\ Ahmad {\it et al.},
                    Phys.\ Rev.\ Lett.\ {\bf 87}, 071301 (2001);
                    Phys.\ Rev.\ Lett.\ {\bf 89}, 011301 (2002);
                    Phys.\ Rev.\ Lett.\ {\bf 89}, 011302 (2002).

\bibitem{SNO2}      SNO Collaboration, S.N.\ Ahmed {\em et al.},
                    Phys.\ Rev.\ Lett.\  {\bf 92}, 181301 (2004).

\bibitem{SNOL}      SNO Collaboration, B.~Aharmim {\it et al.},
                    Phys.\ Rev.\ C {\bf 72}, 055502 (2005).

\bibitem{Kam1}      KamLAND Collaboration, K.~Eguchi {\it et al.},
                    Phys.\ Rev.\ Lett.\  {\bf 90}, 021802 (2003).

\bibitem{Kam2}      KamLAND Collaboration, T.\ Araki {\it et al.},
                    Phys.\ Rev.\ Lett.\  {\bf 94}, 081801 (2005).


\bibitem{Rev}       G.L.\ Fogli, E.\ Lisi, A.\ Marrone, and A.\ Palazzo,
                    Prog.\ Part.\ Nucl.\ Phys.\ {\bf 57}, 742 (2006).

\bibitem{Rev_val}   M.\ Maltoni, T.\ Schwetz, M.A.\ Tortola and J.W.F.\ Valle,
                    New J.\ Phys.\  {\bf 6}, 122 (2004) [arXiv:hep-ph/0405172].


\bibitem{gf04}      G.L.\ Fogli, E.\ Lisi, A.\ Marrone, and A.\ Palazzo,
                    Phys.\ Lett.\ B {\bf 583}, 149 (2004);
                    G.~Fogli and E.~Lisi,
                    New J.\ Phys.\  {\bf 6}, 139 (2004).

\bibitem{Robust}    O.G.\ Miranda, M.A.\ Tortola and J.W.F.\ Valle,
                    JHEP {\bf 0610}, 008 (2006).


\bibitem{PDG}       Particle Data Group, W.M.\ Yao {\it et al.},
                    J.\ Phys.\ G {\bf 33}, 1 (2006).

\bibitem{Ancho}     L.A.\ Anchordoqui, J.\ Phys.\ Conf.\ Ser.\  {\bf 60}, 191 (2007);
                    L.\ Anchordoqui and F.\ Halzen,  Annals Phys.\  {\bf 321}, 2660 (2006).


\bibitem{string}    J.\ Ellis, N.E.\ Mavromatos, D.V.\ Nanopoulos,
                    and E.\ Winstanley, Mod.\ Phys.\ Lett.\ A {\bf 12}, 243 (1997);
                    J.R.\ Ellis, N.E.\ Mavromatos, and D.V.\ Nanopoulos,
                    Mod.\ Phys.\ Lett.\  A {\bf 12}, 1759 (1997);
                    R.\ Gambini, R.A.\ Porto, and J.\ Pullin,
                    Class.\ Quant.\ Grav.\ {\bf 21}, L51 (2004).

\bibitem{Schwetz}   T.\ Schwetz,
                    Phys.\ Lett.\ B {\bf 577}, 120 (2003).

\bibitem{Adiabat}   A.Yu.\ Smirnov, in the Proceedings of IPM School and
                    Conference on Lepton and Hadron Physics (IPM-LHP06),
                    Teheran, Iran, 2006; hep-ph/0702061.


\bibitem{BSOP}      J.N.\ Bahcall, A.M.\ Serenelli and S.\ Basu,
                    Astrophys.\ J.\  {\bf 621}, L85 (2005).


\bibitem{Borex}     Borexino Collaboration, G.~Alimonti {\it et al.}
                    Astropart.\ Phys.\  {\bf 16}, 205 (2002).


\bibitem{SKLE}      Super-Kamiokande Collaboration,
                    Y.~Ashie {\it et al.},
                    Phys.\ Rev.\ Lett.\  {\bf 93}, 101801 (2004).


\bibitem{Lore94}    F.N.\ Loreti and A.B.\ Balantekin,
                    Phys.\ Rev.\ D {\bf 50}, 4762 (1994).


\bibitem{Nuno96}    H.\ Nunokawa, A.\ Rossi, V.B.\ Semikoz, and J.W.F.\ Valle,
                    Nucl.\ Phys.\ B {\bf 472}, 495 (1996).

\bibitem{Bala96}    A.B.\ Balantekin, J.M.\ Fetter, and F.N. Loreti,
                    Phys.\ Rev.\ D {\bf 54}, 3941 (1996).

\bibitem{Burg97}    C.P.\ Burgess and D.\ Michaud,
                    Annals.\ Phys.\ {\bf 256}, 1 (1997).

\bibitem{Byko00}    A.A.\ Bykov, M.C.\ Gonzalez-Garcia, C.\ Pena-Garay, V.Y.\ Popov and V.B.\ Semikoz, hep-ph/0005244.

\bibitem{Burg03}    C.P.\ Burgess, N.S.\ Dzhalilov, M.\ Maltoni, T.I.\ Rashba, V.B.\ Semikoz, M.A.\ Tortola and J.W.F.\ Valle,
                    Astrophys.\ J.\ {\bf 588}, L65 (2003).

\bibitem{Guzz03}    M.M.\ Guzzo, P.C.\ de Holanda, and N.\ Reggiani,
                    Phys.\ Lett.\ B {\bf 569}, 45 (2003); N.\ Reggiani, M.M.\ Guzzo,
                    and P.C.\ de Holanda, Braz.\ J.\ Phys.\ {\bf 34}, 1729 (2004).

\bibitem{Bala03}    A.B.\ Balantekin and H.\ Y\"{u}ksel,
                    Phys.\ Rev.\ D {\bf 68}, 013006 (2003).

\bibitem{Burg04}    C.P.\ Burgess, N.S.\ Dzhalilov, M.\ Maltoni, T.I.\ Rashba, V.B.\ Semikoz, M.A.\ Tortola and J.W.F.\ Valle,
                    JCAP {\bf 0401}, 007 (2004).

\bibitem{Bena04}    F.\ Benatti and R.\ Floreanini,
                    Phys.\ Rev.\ D {\bf 71}, 013003 (2005).

\bibitem{Petc}      S.T.~Petcov and T.~Schwetz,
                    Phys.\ Lett.\  B {\bf 642}, 487 (2006).

\bibitem{Lind76}    G.\ Lindblad, Commun.\ Math.\ Phys.\ {\bf 48}, 119 (1976).

\bibitem{Gori76}    V.\ Gorini, A.\ Frigerio, M.\ Verri, A.\ Kossakowski, and
                    E.C.G.\ Sudarshan,
                    J.\ Math.\ Phys.\ (N.Y.) {\bf 17}, 821 (1976).

\bibitem{Bena88}    F.\ Benatti and H.\ Narnhofer,
                    Lett.\ Math.\ Phys.\ {\bf 15}, 325 (1988).

\bibitem{Jliu}      J.\ Liu, Phys.\ Lett.\ B {\bf 314}, 52 (1993).


%%%%%%%%%%%%%%%%%%%%%%%%%%%%%%%%%%%%%%%%%%%%%%%%%%%%%%%%%%%%%%%%%%%%%%%%%%%%%%%

\end{thebibliography}
\end{document}